\documentclass[aps,prl,twocolumn,superscriptaddress,amsfonts]{revtex4-1}

\usepackage{graphicx}
\usepackage{layouts}
\usepackage{pstool}
\usepackage{amssymb}
\usepackage[squaren]{SIunits}

\begin{document}

\title{Transition from static to kinetic friction: Insights from a 2D model}

\author{J. Tr{\o}mborg}
\affiliation{Physics of Geological Processes, University of Oslo, P.O. Box 1048 Blindern, 0316 Oslo, Norway}
\email[]{jorgen.tromborg@fys.uio.no}
\author{J. Scheibert}
\affiliation{Physics of Geological Processes, University of Oslo, P.O. Box 1048 Blindern, 0316 Oslo, Norway}
\affiliation{Laboratoire de Tribologie et Dynamique des Syst\`emes, CNRS, Ecole Centrale de Lyon, 36 Avenue Guy de Collongue, 69134 Ecully, France}
\author{D. S. Amundsen}
\affiliation{Physics of Geological Processes, University of Oslo, P.O. Box 1048 Blindern, 0316 Oslo, Norway}
\author{K. Th{\o}gersen}
\affiliation{Physics of Geological Processes, University of Oslo, P.O. Box 1048 Blindern, 0316 Oslo, Norway}
\author{A. Malthe-S{\o}renssen}
\affiliation{Physics of Geological Processes, University of Oslo, P.O. Box 1048 Blindern, 0316 Oslo, Norway}

\begin{abstract}
We describe a 2D spring-block model for the transition from static to kinetic friction at an elastic slider/rigid substrate interface obeying a minimalistic friction law (Amontons-Coulomb). By using realistic boundary conditions, a number of previously unexplained experimental results on precursory micro-slip fronts are successfully reproduced. From the analysis of the interfacial stresses, we derive a prediction for the evolution of the precursor length as a function of the applied loads, as well as an approximate relationship between microscopic and macroscopic friction coefficients. We show that the stress build-up due to both elastic loading and micro-slip-related relaxations depend only weakly on the underlying shear crack propagation dynamics. Conversely, crack speed depends strongly on both the instantaneous stresses and the friction coefficients, through a non-trivial scaling parameter.
\end{abstract}

\maketitle

Frictional interfaces are important in many areas of science and technology, including seismology \cite{Scholz-CUP-2002}, biology \cite{Urbakh-Klafter-Gourdon-Israelachvili-Nature-2004, Scheibert-Leurent-Prevost-Debregeas-Science-2009} and nanomechanics \cite{Bhushan-Springer-2008}. Whereas a satisfactory picture of the steady sliding regime of such interfaces has been developed during the last twenty years \cite{Persson-Springer-2000, Baumberger-Caroli-AdvPhys-2006, Scheibert-Prevost-Debregeas-Katzav-AddaBedia-JMechPhysSolids-2009}, the dynamics of the transition from static to kinetic friction remains elusive. During the last decade, a renewed interest has grown in such transitions, due to experimental studies that directly measured the local dynamics of frictional interfaces \cite{Baumberger-Caroli-Ronsin-PhysRevLett-2002, Xia-Rosakis-Kanamori-Science-2004, Rubinstein-Cohen-Fineberg-Nature-2004, Scheibert-Debregeas-Prevost-CondMat-2008, Chateauminois-Fretigny-Olanier-PhysRevE-2010}. They have shown that macroscopic sliding occurs only after shear crack-like micro-slip fronts have spanned the entire contact interface.

\textit{Experimentally}, micro-slip front nucleation, propagation and arrest was shown to be controlled by the instantaneous stress field at the interface. Fronts nucleate preferentially at the trailing edge of the contact area \cite{Baumberger-Caroli-Ronsin-PhysRevLett-2002, Rubinstein-Cohen-Fineberg-Nature-2004, Rubinstein-Cohen-Fineberg-PhysRevLett-2007, Bennewitz-etal-JPhysCondensMatter-2008, Maegawa-Suzuki-Nakano-TribolLett-2010, BenDavid-Cohen-Fineberg-Science-2010}, an effect explained either by the enhanced shear stress near the loading point in side-driven systems \cite{Rubinstein-Cohen-Fineberg-Nature-2004, Rubinstein-Cohen-Fineberg-PhysRevLett-2007, Braun-Barel-Urbakh-PhysRevLett-2009, Maegawa-Suzuki-Nakano-TribolLett-2010} or by a friction-induced pressure asymmetry in top-driven systems \cite{Baumberger-Caroli-Ronsin-PhysRevLett-2002, Scheibert-Dysthe-EPL-2010}. Fronts can arise well below the macroscopic static friction threshold and arrest before the whole contact area has ruptured \cite{Rubinstein-Cohen-Fineberg-PhysRevLett-2007, Bennewitz-etal-JPhysCondensMatter-2008, Maegawa-Suzuki-Nakano-TribolLett-2010}. The length and number of these \textit{precursors} depends on the precise way in which shear \cite{Rubinstein-Cohen-Fineberg-PhysRevLett-2007} and normal \cite{Maegawa-Suzuki-Nakano-TribolLett-2010} forces are applied. Moreover, precursors are associated with significant changes in the spatial distribution of the real contact area \cite{Rubinstein-Cohen-Fineberg-PhysRevLett-2007}, a quantity related to the local interfacial pressure. Finally, the propagation speed of micro-slip fronts, which covers a wide range \cite{Baumberger-Caroli-Ronsin-PhysRevLett-2002, Xia-Rosakis-Kanamori-Science-2004, Rubinstein-Cohen-Fineberg-Nature-2004, Nielsen-Taddeucci-Vinciguerra-GeophysJInt-2010}, correlates with the local shear to normal stress ratio at nucleation \cite{BenDavid-Cohen-Fineberg-Science-2010}.

\textit{Theoretically}, some aspects of these observations have been studied using one-dimensional (1D) models. The conditions leading to a large range of front velocities were addressed using a 1D spring-block model with a time-dependent friction law \cite{Braun-Barel-Urbakh-PhysRevLett-2009}. The role of an asymmetric normal loading on the length of precursors was considered using a 1D spring-block model with Amontons-Coulomb (A-C) friction and different normal forces ascribed to different blocks \cite{Maegawa-Suzuki-Nakano-TribolLett-2010}. The dependence of the series of precursors on the friction-induced pressure asymmetry was described, for A-C friction, using a quasi-static 1D model \cite{Scheibert-Dysthe-EPL-2010}. In these three studies, the normal stress distribution was assumed to be uniform \cite{Braun-Barel-Urbakh-PhysRevLett-2009} or linear (either fixed \cite{Maegawa-Suzuki-Nakano-TribolLett-2010} or friction-dependent \cite{Scheibert-Dysthe-EPL-2010}). Such assumptions impede quantitative comparison with experiments since determination of the actual stress field requires solution of the elastic problem for the two bodies in contact, including their precise geometry, elastic properties and boundary conditions, not only at the frictional interface but also on all their other boundaries. As a first step towards such complete description, we present a minimal 1+1D (along + orthogonal to interface) model for the transition from static to kinetic friction of an elastic slider on a rigid substrate. We show that this model, by enabling realistic boundary conditions, is sufficient to reproduce a series of still unexplained experimental observations.

\begin{figure}[b!]
\resizebox{.49\columnwidth}{!}{\includegraphics{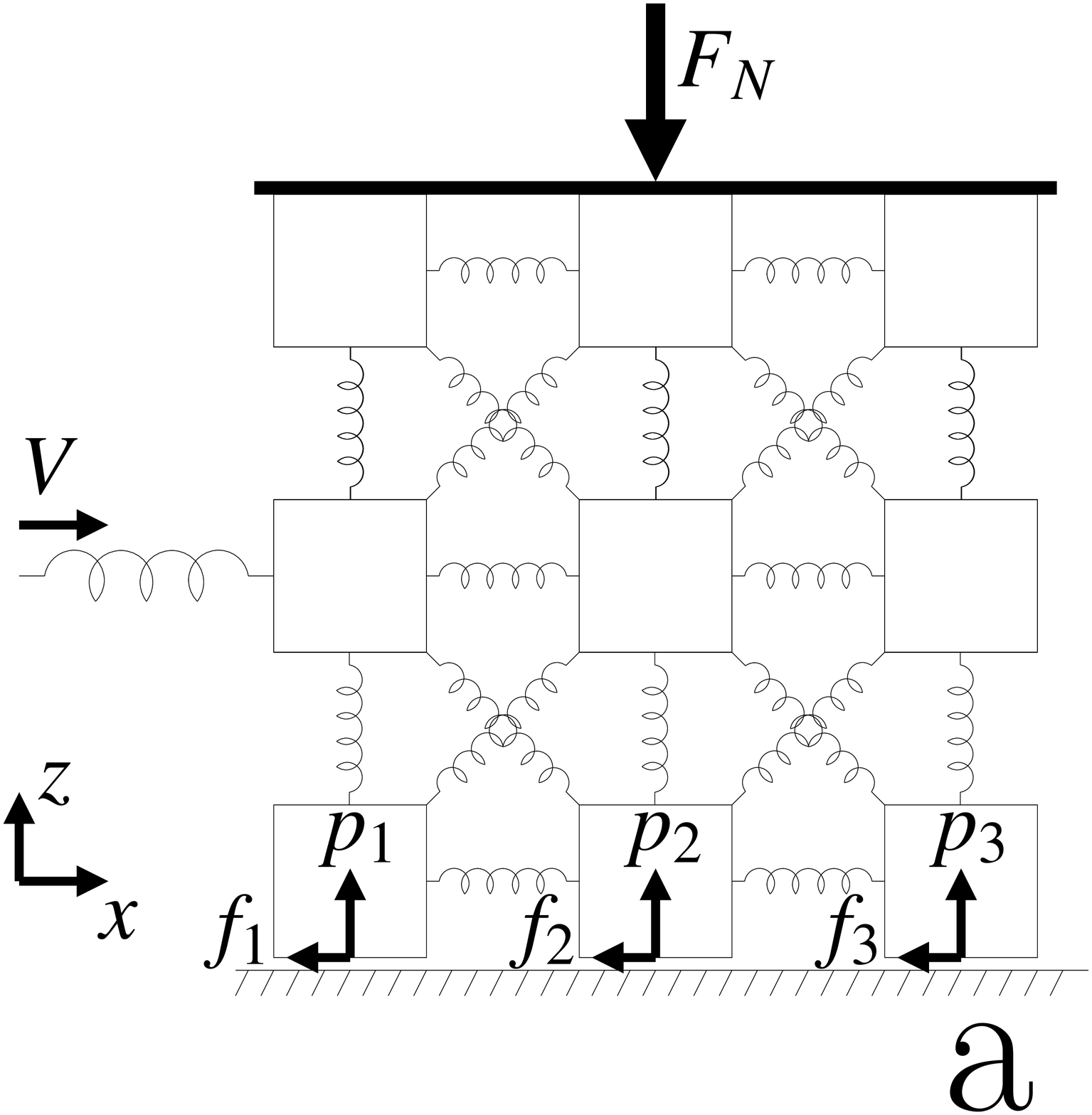}}
\resizebox{.49\columnwidth}{!}{\includegraphics{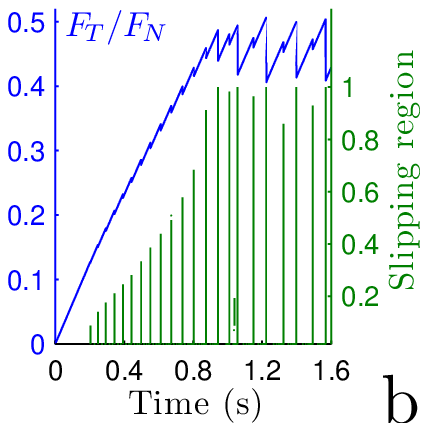}}
\caption{(a): Sketch of the 2D spring-block model. (b) Blue: Typical loading curve, from static contact to macroscopic stick-slip, for setup 1 ($h$=5$\SIunits{\milli\meter}$, $L$=140$\SIunits{\milli\meter}$, $F_N$=3500$\SIunits{\newton}$). Green: Locations of slipping regions for all micro-slip event. 0 (1) corresponds to the trailing (leading) edge.
\label{fig1}}
\end{figure}

We used the 2D spring-block model sketched in Fig. \ref{fig1}(a). The slider has mass $M$ and sizes $L$ and $H$ in the horizontal ($x$) and vertical ($z$) directions, respectively. It is divided along a square lattice into $N = N_x N_z$ blocks of mass $m= M/N$. Blocks are coupled to their four nearest neighbours and their four next-nearest neighbours by springs of equilibrium lengths $l=L/(N_x-1)=H/(N_z-1)$ and $\sqrt{2}l$ and stiffnesses $k$ and $k/2$, respectively, giving an isotropic elastic model with Poisson's ratio $1/3$. The force exerted on block $i$ by block $j$ is thus $k_{ij} (r_{ij}-l_{ij}) \frac{\Delta \textbf{x}_{ij}}{r_{ij}}$ when blocks are connected, 0 otherwise, where $\textbf{x}=(x,z)$, $\Delta \textbf{x}_{ij} = \textbf{x}_j-\textbf{x}_i$, $r_{ij} = \left|\Delta \textbf{x}_{ij}\right|$ and $k_{ij}$ and $l_{ij}$ are the stiffness and equilibrium length of the spring connecting blocks $i$ and $j$. Block oscillations are damped using a viscous force $\eta(\dot{\textbf{x}}_{j}-\dot{\textbf{x}}_{i})$ on the relative motion of connected blocks. We chose the coefficient $\eta=\sqrt{0.1 k m}$ so that blocks are underdamped and event-triggered oscillations die out well before the next event. All $\eta$ satisfying these conditions gave similar results.

Boundary conditions were designed to model two different experiments described in the literature: \cite{Rubinstein-Cohen-Fineberg-PhysRevLett-2007} (setup 1) and \cite{Maegawa-Suzuki-Nakano-TribolLett-2010} (setup 2). They differ in the way the top blocks are loaded. For setup 1, they are glued to a rigid rod of mass 75.6$\SIunits{\gram}$, itself submitted to a normal force $F_N$ and coupled to a soft "spring mattress" of stiffness 0.4$\SIunits{\mega\newton\per\meter}$, the effect of which is modeled with a restoring torque proportional to the rod's tilt angle. For setup 2, the top blocks are submitted to a linear time-independent distribution of vertical forces $\frac{F_N}{N_x} \left(1+ \frac{2i-N_x-1}{N_x-1}\theta\right)$, where $\theta\in [-1,1]$ controls the pressure asymmetry. In both setups, the bottom blocks lie on an elastic foundation of modulus $k_f=k$, i.e. each block is submitted to a vertical force of amplitude $p_i=k_f \left|z_i\right|$ if $z_i<0$ or 0 otherwise, where $z_i$ is the vertical displacement of block $i$. All $k_f>k$ gave similar results, so that the substrate can be considered rigid compared to the slider. Both vertical boundaries are free, except for a horizontal driving force $F_T=K(Vt-x_h)$ applied on the left-side block situated at height $h$ above the interface, where $x_h$ is the $x$-displacement of this block. This models a pushing device of stiffness $K$ driven at a small constant velocity $V$. The amplitudes $f_i$ of the friction forces on the bottom blocks follow the minimalistic local A-C friction law with static and kinetic friction coefficients $\mu_s$ and $\mu_k<\mu_s$. If $\dot{x}_{i}=0$, $f_i$ balances all horizontal forces on block $i$ up to $\mu_s p_i$, at which slip initiates; then $f_i=\pm \mu_k p_i$ when $\dot{x}_{i} \lessgtr 0$. The $2N$ equations of motion are solved simultaneously using a fourth order Runge--Kutta integrator on a uniform temporal grid of resolution $\Delta t$.

\begin{figure}
\resizebox{.5\columnwidth}{!}{\includegraphics{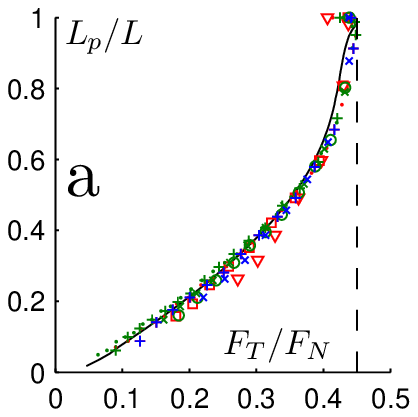}}%
\resizebox{.5\columnwidth}{!}{\includegraphics{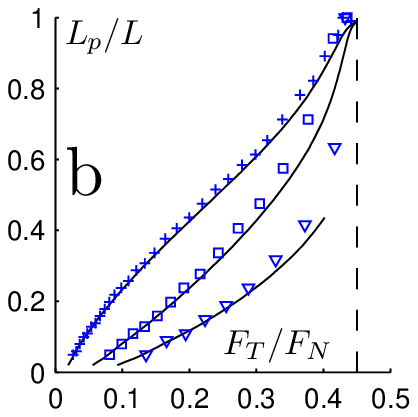}}%
\caption{Normalized precursor length $L_p/L$ as a function of normalized shear force $F_T / F_N$. (a) Setup 1 for various configurations: $L$=140 or 200$\SIunits{\milli\meter}$, $h$ ranges between 2.5 and 15$\SIunits{\milli\meter}$, $F_N$=1750, 2700 or 3500$\SIunits{\newton}$, $H$=75$\SIunits{\milli\meter}$, $K$=4$\SIunits{\mega\newton\per\meter}$, $M$=75.6$\SIunits{\gram}$, $V$=0.7875$\SIunits{\milli\meter\per\second}$, $l$=2.5$\SIunits{\milli\meter}$, $k$=13.5$\SIunits{\mega\newton\per\meter}$, $\mu_s$=0.7, $\mu_k$=0.45, $\Delta t$=0.2$\SIunits{\micro\second}$. (b) Setup 2 for $\theta$=0.833 ($+$), 0 ($\square$) or -0.833 ($\triangledown$), $L$=100$\SIunits{\milli\meter}$, $H$=20$\SIunits{\milli\meter}$, $K$=0.8$\SIunits{\mega\newton\per\meter}$, $M$=12$\SIunits{\gram}$, $V$=0.45$\SIunits{\milli\meter\per\second}$, $F_N$=400$\SIunits{\newton}$, $h$=2$\SIunits{\milli\meter}$, $l$=1$\SIunits{\milli\meter}$, $k$=9.375$\SIunits{\mega\newton\per\meter}$, $\mu_s$=0.7, $\mu_k$=0.45, $\Delta t$=0.09$\SIunits{\micro\second}$. Vertical dashes: $\mu_k$. Solid lines: prediction obtained with $\mu_s = \mu_k$.
\label{fig2}}
\end{figure}

Using this model, we simulated the transition from static to kinetic friction for the various loading configurations reported in \cite{Rubinstein-Cohen-Fineberg-PhysRevLett-2007} and \cite{Maegawa-Suzuki-Nakano-TribolLett-2010}. Model parameters, chosen in accordance with experiments \footnote{$k=\frac{3BE}{4}$. $B$ ($E$) is the slider thickness (Young modulus)}, are given in the caption of Fig. \ref{fig2}. Figure \ref{fig1}(b) (blue) shows a typical loading curve for setup 1, from static contact up to macroscopic stick-slip. As in 1D models, and in agreement with experimental results, we observe, well before macroscopic sliding, a series of partial force relaxation events, corresponding to precursors, which all nucleate at the trailing edge, and extend over increasing lengths $L_p$ (Fig. \ref{fig1}(b), green). 

We first focus on the dependence of $L_p$ on the applied tangential force $F_T$ just after relaxation. We simulated, for setup 1, different slider lengths $L$, pushing heights $h$ and normal forces $F_N$. The behaviour under different conditions differ by the number of precursors occurring along the transition: the increase in both $L_p$ and $F_T$ between precursors scales almost linearly with $h$ for small $h$, so that larger $L$ and smaller $h$ yield more precursors. However, Fig. \ref{fig2}(a) shows that the results for all conditions can be collapsed on a single curve by plotting $L_p/L$ as a function of $F_T/F_N$. The same collapse was found in \cite{Rubinstein-Cohen-Fineberg-PhysRevLett-2007}, with a very similar shape for the non-linear increase of $L_p/L$ with $F_T/F_N$. In particular, we reproduce the transition from a roughly linear increase up to $L_p / L \sim$0.5 to a more rapid growth for longer precursors. We emphasize that a 1D model with homogeneous normal loading would produce a purely linear increase. In Fig. \ref{fig2}(b) we compare, for setup 2, the evolutions of $L_p/L$ as a function of $F_T/F_N$ for three different linearly asymmetric normal loadings of the slider. Qualitatively, the lower the normal load on the trailing edge, the lower the threshold force required to nucleate precursors, and therefore the lower the tangential force $F_T$ necessary for the precursor to reach a given length $L_p$, which explains the relative positions of the three curves in Fig. \ref{fig2}(b). Again, the simulated curves are in excellent agreement with the experimental results in Fig. 8 in \cite{Maegawa-Suzuki-Nakano-TribolLett-2010}, much better than the 1D simulation (Fig. 13 in \cite{Maegawa-Suzuki-Nakano-TribolLett-2010}). These non-trivial results for both setups clearly demonstrate that, by enabling realistic boundary conditions, 2D models do offer significantly improved agreement with experiments. 

\begin{figure}
\resizebox{.5\columnwidth}{!}{\includegraphics{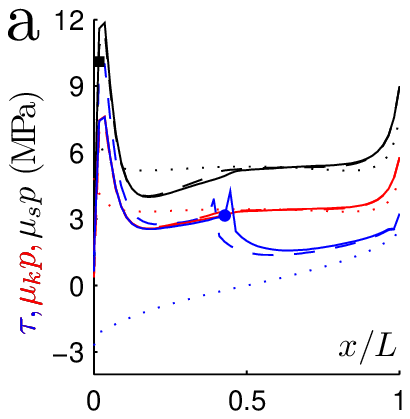}}%
\resizebox{.5\columnwidth}{!}{\includegraphics{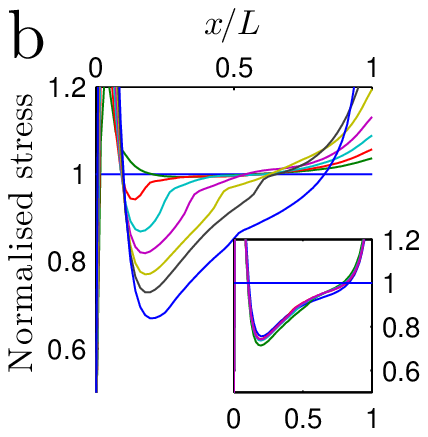}}%
\caption{Spatial distribution of stresses at three stages of the transition. Dotted lines: initial configuration ($F_T$=0). Dashed (solid) lines: at crack initiation (arrest) for the 9th precursor shown in Fig. \ref{fig1}(b). Blue: $\tau$. Black: $\mu_s p$. Red: $\mu_k p$. Square (disk): nucleation (arrest) point. (b) Spatial distribution of $p$, normalized by the initial distribution, after each of the precursors shown in Fig. \ref{fig1}(b). Inset: Similar distribution for system-sized events during macroscopic stick-slip.
\label{fig3}}
\end{figure}

We now use our simulation results to gain insight into how shear cracks nucleate, propagate and arrest at an interface obeying A-C friction. In particular, we will examine the role played by the evolution of the shear and normal stress distributions at the interface. We emphasize that, in 2D, the latter is a result of the simulation, not an assumption, which is required in 1D. In Fig. \ref{fig3}(a) we show the typical stress evolution for setup 1. Initially ($F_T=0$) the normal stress $p(x)$ is symmetric, with edge effects related to the flat punch geometry of the contact. The shear stress $\tau(x)$ is antisymmetric, due to friction-frustrated Poisson expansion. These stresses are in excellent agreement with those expected from contact mechanics and those measured in \cite{BenDavid-Cohen-Fineberg-Science-2010}. Application of a tangential force at the trailing edge modifies slightly the normal stress field and modifies significantly the shear stress field, with a large increase near the trailing edge over a distance of order $h$. The local slipping threshold $\tau(x) = \mu_s p(x)$ is therefore reached first near the trailing edge and a micro-slip front corresponding to the first precursor nucleates there. It then stops after propagation over a finite distance $L_p$ and the whole scenario is repeated until the leading edge is reached. Figure \ref{fig3}(a) illustrates this scenario for a typical precursor event. 

We find that the shear stress $\tau (x)$ just after a precursor is always very close to $\mu_k p (x)$ over the whole slipped length $x \leq L_p$. This shows that the arrest state of the interface is only weakly dependent of the static friction coefficient $\mu_s$, and is controlled primarily by the kinetic friction coefficient $\mu_k$. We emphasize that this behaviour is not specific to A-C friction, but remained true for slip-weakening friction, provided the weakening distance is smaller than a few $\mu$m. Based on this robust behaviour, we propose the following procedure to predict, for given loading conditions and a given $\mu_k$, the non-linear evolution of $L_p$ with $F_T$: We run the model for equal friction coefficients ($\mu_s=\mu_k$). The transition to kinetic friction in this simplified model is smooth, with a continuously growing micro-slip region the length of which is shown as a function of the applied force in Fig. \ref{fig2} (solid lines). For both setups and for all loading configurations, this curve is in good quantitative agreement with the curve for the length of precursors as a function of the arrest force. Such agreement is due to the fact that any arrest state in the full model is very similar to the state reached in the simplified model for the same force $F_T$: (i) in the micro-slip region $\tau(x)=\mu_k p(x)$ and (ii) in the stuck region $\tau(x)$ arises from almost identical boundary conditions. Because internal dynamics are fast compared to changes in external loading, the prediction is $V$-independent and can be obtained using simple equilibrium calculations, i.e. it does not require a complete dynamical simulation.

The very last precursor in the full model propagated over almost the entire interface and left a shear stress that was equal almost everywhere to $\mu_k p$, yielding a total shear force $F_T \simeq \mu_k F_N$. The last increment of tangential force required to trigger the first system-sized event brings the vicinity of the trailing edge to its threshold, whereas the shear stress on the rest of the interface is essentially unchanged, yielding a maximum total shear force only slightly above $\mu_k F_N$ (see Fig. \ref{fig1}(b)). This maximum force is classically interpreted as $\mu_s^\mathrm{macro} F_N$ with $\mu_s^\mathrm{macro}$ the macroscopic static friction coefficient. Therefore our results suggest that, in side-driven systems, whatever the value of $\mu_s$, $\mu_s^\mathrm{macro} \simeq \mu_k$. The difference between the macro- and microscopic static friction coefficients, already discussed in recent 1D models \cite{Maegawa-Suzuki-Nakano-TribolLett-2010, Scheibert-Dysthe-EPL-2010}, provides a possible explanation for the anomalously high values of $\mu_s$ compared to $\mu_s^\mathrm{macro}$ reported in \cite{BenDavid-Cohen-Fineberg-Science-2010}.

Not only the shear stress $\tau$ but also the normal stress $p$ is altered along the transition. Figure \ref{fig3}(b) shows $p(x)$, normalized by the initial distribution (at $F_T=0$), after each of the successive precursors. The normal stress is found to be significantly decreased along the path of the precursor that just occurred, by up to 30$\%$ whereas, apart from edge effects, it is mainly unchanged in front of it. The normal stress then assumes a reproducible distribution in the macroscopic stick-slip regime (Fig. \ref{fig3}(b), inset). Recalling that, at normally loaded rough contact interfaces, normal stress is robustly found to be locally proportional to the area of real contact, the results of Fig. \ref{fig3}(b) show very good agreement with measurements of the real area of contact reported in \cite{Rubinstein-Cohen-Fineberg-PhysRevLett-2007} (Fig. 4a of \cite{Rubinstein-Cohen-Fineberg-PhysRevLett-2007}). Direct quantitative comparison is not possible, mainly because normal stress has a constant integral (normal force $F_N$ imposed) whereas the total real area of contact is not a conserved quantity, but typically decreases by 20$\%$ across the transition \cite{Rubinstein-Cohen-Fineberg-PhysRevLett-2007, BenDavid-Rubinstein-Fineberg-Nature-2010}.

\begin{figure}
\resizebox{.5\columnwidth}{!}{\includegraphics{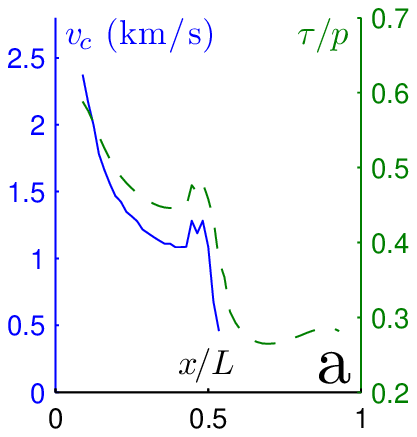}}%
\resizebox{.5\columnwidth}{!}{\includegraphics{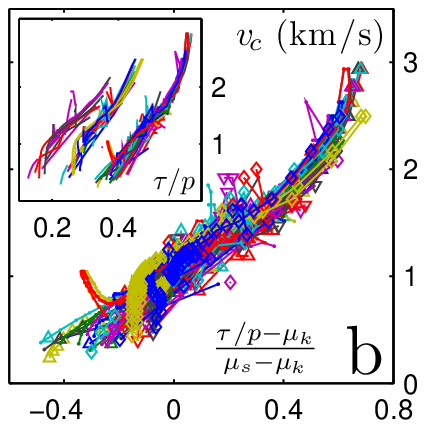}}%
\caption{(a) Local crack speed $v_c$ (solid line, spatial average on 5 neighbouring blocks) and local shear to normal stress ratio $\tau / p$ (dashed line) as a function of crack length $x$ for the 11th precursor event shown in Fig. \ref{fig1}(b). (b) Inset: $v_c$ as a function of $\tau / p$. Runs: 1 (left), 2 (center), 3 and 4 (right). Main: $v_c$ as a function of $\frac{\tau / p -\mu_k}{\mu_s - \mu_k}$. Runs: 1 ($\triangledown$), 2 ($\diamond$), 3 ($\triangle$), 4 ($\cdot$). The longitudinal (shear) wave speed is 1677m/s (968m/s).\\
Run 1: $L$=140$\SIunits{\milli\meter}$, $F_N$=3500$\SIunits{\newton}$, $h$=5$\SIunits{\milli\meter}$, $\mu_s$=0.5, $\mu_k$=0.2. 
Run 2: $L$=140$\SIunits{\milli\meter}$, $F_N$=3500$\SIunits{\newton}$, $h$=5$\SIunits{\milli\meter}$, $\mu_s$=0.55, $\mu_k$=0.3. 
Run 3: $L$=140$\SIunits{\milli\meter}$, $F_N$=3500$\SIunits{\newton}$, $h$=7.5$\SIunits{\milli\meter}$, $\mu_s$=0.7, $\mu_k$=0.45. 
Run 4: $L$=200$\SIunits{\milli\meter}$, $F_N$=2700$\SIunits{\newton}$, $h$=10$\SIunits{\milli\meter}$, $\mu_s$=0.7, $\mu_k$=0.45.
\label{fig4}}
\end{figure}

Let us now focus on the propagation dynamics of the simulated frictional shear cracks. Figure \ref{fig4}(a) shows the evolution of the local crack speed $v_c$ as a function of position $x$ along the interface for a typical precursor event in setup 1. $v_c(x)$ decreases from around 2400$\SIunits{\meter\per\second}$ near the trailing edge to around 400$\SIunits{\meter\per\second}$ just before crack arrest and appears to be strongly correlated to the shear to normal stress ratio $\tau (x) / p(x)$ (Fig. \ref{fig4}(a)). In fact, when $v_c$ is plotted as a function of $\tau / p$, all points corresponding to all locations $x$ along all successive events in a given simulation collapse on a single curve (Fig. \ref{fig4}(b), inset). Different geometries or loading conditions yield the very same curve but different friction coefficients yield different curves (Fig. \ref{fig4}(b), inset). All these curves can then be collapsed on the same master curve when $v_c(x)$ is plotted as a function of the non-trivial parameter $\frac{\tau(x) / p(x) -\mu_k}{\mu_s - \mu_k}$, which represents the local distance to the slipping threshold and extends existing parameters \cite{Muratov-PhysRevE-1999, Scholz-CUP-2002} to spatial heterogenities in normal stress $p(x)$. These results are a generalization, accounting for any value of the microscopic friction coefficients, of a similar collapse obtained for experimental data, from system-sized events only, using $\tau  / p$ as a parameter (Fig. 3 in \cite{BenDavid-Cohen-Fineberg-Science-2010}). Our results also suggest that, if probed, precursors would follow the same experimental curve as system-sized events.

In surprising contrast with the excellent agreement found with experiments up to now, the master curve of Fig. \ref{fig4}(b) exhibits strong discrepancies with its experimental counterpart. The shape as well as the explored ranges of both $v_c$ and $\tau / p$ are different. In particular, our model does not produce very slow micro-slip fronts like those observed in various experiments \cite{Nielsen-Taddeucci-Vinciguerra-GeophysJInt-2010, BenDavid-Cohen-Fineberg-Science-2010}. Most likely this is because the minimalistic friction used here (A-C) lacks some time-dependent ingredient necessary to yield slow fronts, like those in \cite{Braun-Barel-Urbakh-PhysRevLett-2009}.

All our results suggest two distinct levels of description of the transition from static to kinetic friction. First, a \textit{kinematic} description of (i) the slow evolution of interfacial stresses between events and (ii) the stress conditions at crack nucleation and arrest. Second, a \textit{dynamic} description of the fast propagation of micro-slip fronts along the interface. Crack dynamics appear to depend crucially, via the friction law, on the kinematic stresses at crack initiation. Conversely, the kinematic description was found essentially independent of the underlying dynamics: we could successfully reproduce all available experimental kinematic results, even with an unrealistic friction law. The key observation yielding this surprising success is that, at crack arrest, $\tau(x)=\mu_k p(x)$ over the whole slipped region. Since the other 2D boundary conditions on the slider could be accurately taken into account, our results strongly suggest that a very similar arrest condition holds in the experiments. We thus believe that any friction law leading to such arrest state will produce as good kinematic agreement as A-C friction, irrespective of the dynamic way of reaching this arrest state.

\begin{acknowledgments}
We thank J. L. Vinningland and M. Dabrowski for discussions, and G. Debr\'egeas and P. Meakin for comments on the manuscript. We acknowledge funding from the European Union (Marie Curie grant PIEF-GA-2009-237089). This paper was supported by a Center of Excellence grant to PGP from the Norwegian Research Council.
\end{acknowledgments}

\end{document}